  \documentclass{PoS}
  
  \usepackage{graphicx}
  \usepackage{subfigure}

  \title{Is an obscured AGN at the centre of the disk galaxy IC~2497 responsible for Hanny's Voorwerp?}

  \ShortTitle{Hanny's Voorwerp}
  
  \author{\speaker{H. Rampadarath }$^{ac}$\thanks{Currently at The School of Physics and Astronomy, University of Manchester, UK} , M.A. Garrett $^{bce}$, T. Muxlow$^{d}$, G. I. G. J\'ozsa$^{b}$, T. A. Oosterloo$^{bf}$, and  Z. Paragi$^{ag}$  \\
 \llap{$^a$} Joint Institute for VLBI in Europe (JIVE), The Netherlands\\
 \llap{$^b$}Netherlands Institute for Radio Astronomy (ASTRON), The Netherlands\\
 \llap{$^c$}Sterrewacht Leiden, Leiden University, The Netherlands\\
 \llap{$^d$}Jodrell Bank Centre for Astrophysics, The University of Manchester, UK\\
 \llap{$^e$}Centre for Supercomputing, Swinburne University of Technology, Australia\\
\llap{$^f$}Kapteyn Astronomical Institute, University Groningen, The Netherlands\\
 \llap{$^g$}MTA Research Group for Physical Geodesy and Geodynamics, Hungary\\
  
  E-mail: \email{rampadarath@jive.nl}, \email{garrett@astron.nl}}
  
  \bigskip

  \abstract {We present the results of VLBI and MERLIN observations of the massive disk galaxy IC~2497.\
  Optical observations of IC 2497 revealed the existence of a giant emission nebula "Hanny's Voorwerp" in the proximity of the galaxy. Earlier short-track 18 cm observations with e-VLBI at $\lambda$ 18~cm, detected a compact radio component (C1) at the centre of IC~2497. The brightness temperature of C1 was measured to be $>$  4 $\times$ 10$^5$ K. Deeper, long-track e-VLBI observations presented here, re-confirm the existence of C1 but also reveal the existence of a second compact component (C2) located about 230 milliarcseconds to the North-East of C1. The brightness temperature of C2 is measured to be $>$  1.4 $\times$ 10$^5$ K, suggesting that both components may be related to AGN activity (e.g. a radio core and jet hotspot). Lower resolution $\lambda$ 18~cm MERLIN observations show both components. C1 is shown to be compact with a slight elongation along the direction of Hanny's Voorwerp, and C2 shows a lot of extended emission in an almost perpendicular direction to the direction of the Voorwerp. Our results continue to support the hypothesis that IC 2497 contains an Active Galactic Nucleus (AGN), and that a jet associated with this AGN clears a path that permits ionising radiation from the AGN to directly illuminate the emission nebula. }
 
  \FullConference{Science and Technology of Long Baseline Real-Time Interferometry:\\
      The 8th International e-VLBI Workshop, EXPReS09\\
      June 22 - 26 2009\\
      Madrid, Spain}

\begin{document}

\section{Introduction}

In early 2008, Dutch school teacher, Hanny van Arkel, discovered one of the most bizarre objects to come out of the GalaxyZoo.org  morphological census \cite{lintott_2009}, SDSS J094103.80 \\+344334.2. This object, now known as "Hanny's Voorwerp" \footnote{'Voorwerp' is the Dutch word for "object"} \cite{lintott_2009} appears as an irregular cloud located $15 - 25\,\rm kpc$ to the South-East of the massive disk galaxy IC~2497 \cite{lintott_2009, josh_2009}. Optical observations indicate a high-ionisation state of the cloud despite the lack of a stellar counterpart \cite{lintott_2009}. The quiescent kinematics as derived from optical spectra suggest ionisation from photons as the predominant ionisation process, rather than ionisation from shocks \cite{lintott_2009, josh_2009}. 

Recent radio observations using the Westerbork Synthesis Telescope (WSRT) \cite{josh_2009}, has shown the existence of a radio continuum source at the central position of IC~2497, with an extension in the direction of Hanny's Voorwerp. In addition, neutral hydrogen is detected around the galaxy and the Voorwerp is probably part of this large gas reservoir. HI is also detected in absorption towards the central radio core. Clearly, obscuring material in the direction of the core of IC 2497 is present, while the extended continuum points towards the presence of a radio jet.  

Here we present results of new e-VLBI and MERLIN observations of IC 2497 to investigate the small-scale structure of the core of IC 2497.

\section{Observations and Initial Results}
\medskip 
\subsection{e-VLBI observations}

IC~2497 is a disk galaxy with a redshift of z = 0.050221 \cite{fisher_iras_1995} at R.A. = 09h41m04.094s and Dec. = +34d43m58.03s (J2000 coordinates extracted from VLA FIRST Survey \cite{becker_first_1995}).

IC~2497 was observed for two hours with the European VLBI Network (EVN) using the e-VLBI technique at $\lambda$ 18~cm, with a data rate of 512 Mbps on the 30th September 2008, and for 10 hours on the 19th May 2009 at $\lambda$ 18~cm with a data rate of 1Gbps . The first observation was taken as part of a multi-wavelength radio survey of IC~2497 and Hanny's Voorwerp \cite{josh_2009}.  In both observations the target was phase-referenced to J0945+3534, a calibrator 1.3 degrees away, figures 1a and 1b show the UV coverage for both observations. The calibration solutions derived from J0945+3534 (including phase and amplitude corrections obtained by hybrid mapping the source) were applied to the IC~2497 data. 

A compact radio source (hereafter C1) was detected from the phase calibrated data set of the first short e-VLBI observation, using natural weighting (Figure 2a). The observed position of this compact radio source (RA 09 41 04.0872 $\pm$ 0.0001 and Dec. 34 43 57.7776 $\pm$ 0.0014, J2000) is offset by approximately 230 milliarcseconds to the South-West of the VLA FIRST position. The output from the \textsc{aips} task \textsc{imfit} suggests an image size of $<$ 60 milliarcseconds and a flux density of  $S_{\rm 1.65\,GHz, VLBI} = 1.09\,\pm\, 0.14\,\rm mJy$ . From this we measure a minimum brightness temperature of $\sim4 \times$ 10$^5$ K.

For the second much longer, e-VLBI observation, the target was phase referenced and IC~2497 was then phase self-calibrated with a long solution interval of 10 minutes averaging across the entire 128 MHz band. A weighted image of robust = 2 is shown in Figure 2b. The image confirms the existence of C1, but it also reveals the existence of another compact component to the North-East of C1. The second radio source (hereafter C2) has a flux density of $S_{\rm 1.65\,GHz, VLBI} = 0.594\,\pm\, 0.004\,\rm mJy$, and for C1 we measure a flux density of  $S_{\rm 1.65\,GHz, VLBI} = 1.020\,\pm\, 0.003\,\rm mJy$. For C2, we determine a minimum brightness temperature of 1.4 $\times$ 10$^5$ K.

\medskip
\subsection{MERLIN observations}

On the 2nd of February 2009,  IC~2497 was further observed at $\lambda$ 18~cm with the Multi-Element Radio Linked Interferometer Network (MERLIN). The MERLIN data was phase-calibrated to 0942+358B, and the phase solutions derived were transferred to the target. No self-calibration techniques were applied as the source was very weak and heavily resolved on the Cambridge-Defford baseline. The naturally weighted image is shown in Figure 3. The lower resolution MERLIN image shows a more complex radio morphology: both C1 and C2 are embedded in extended emission that flares out along the direction to the Voorwerp and also perpendicular to this direction.

\bigskip
\begin{figure}
\centering
\begin{tabular}{cc}

\begin{minipage}{2.5in}
\centering
\includegraphics[scale=0.35]{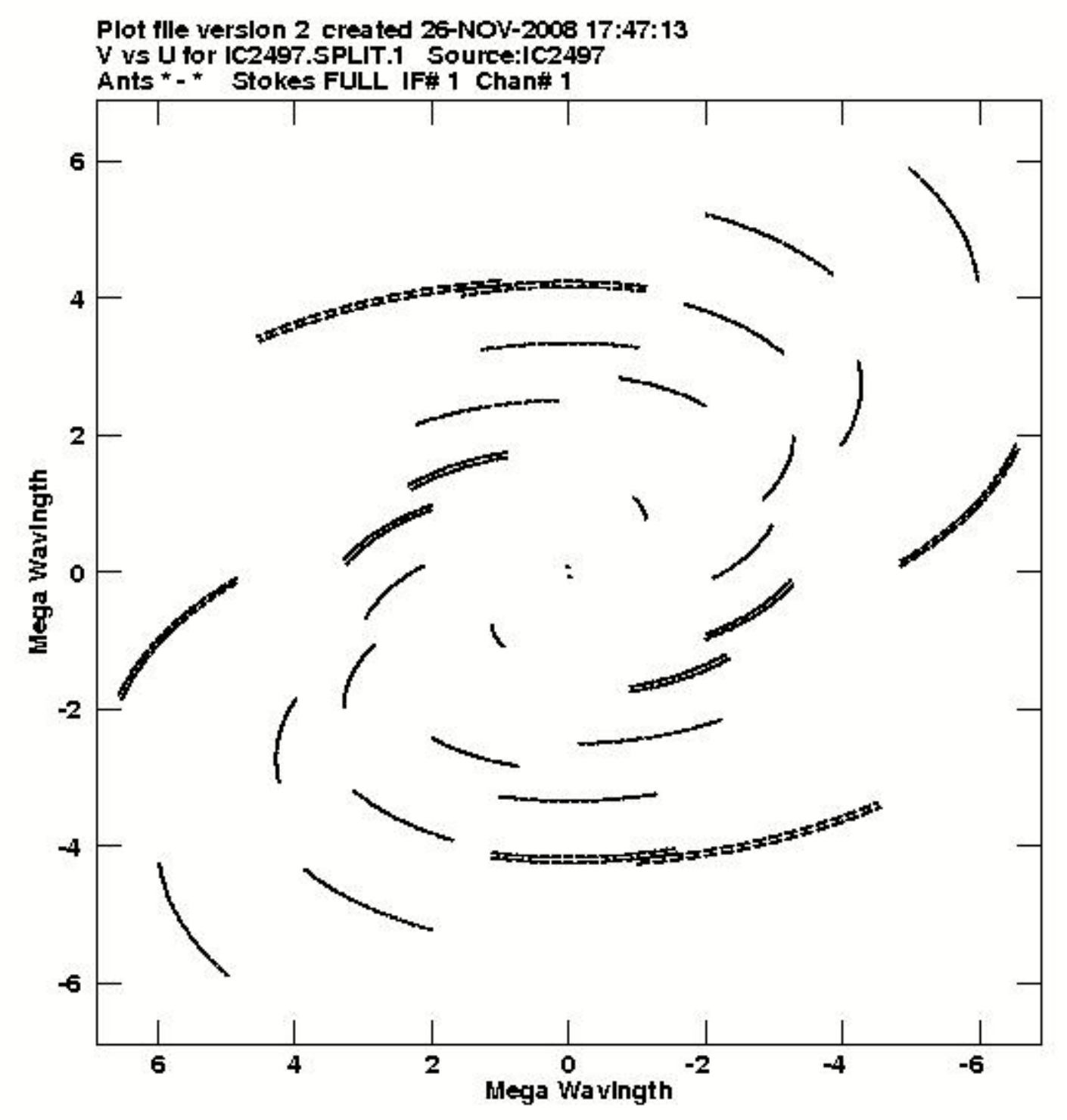} 
\end{minipage}
&
\begin{minipage}{2.5in}
\centering
\includegraphics[scale=0.35]{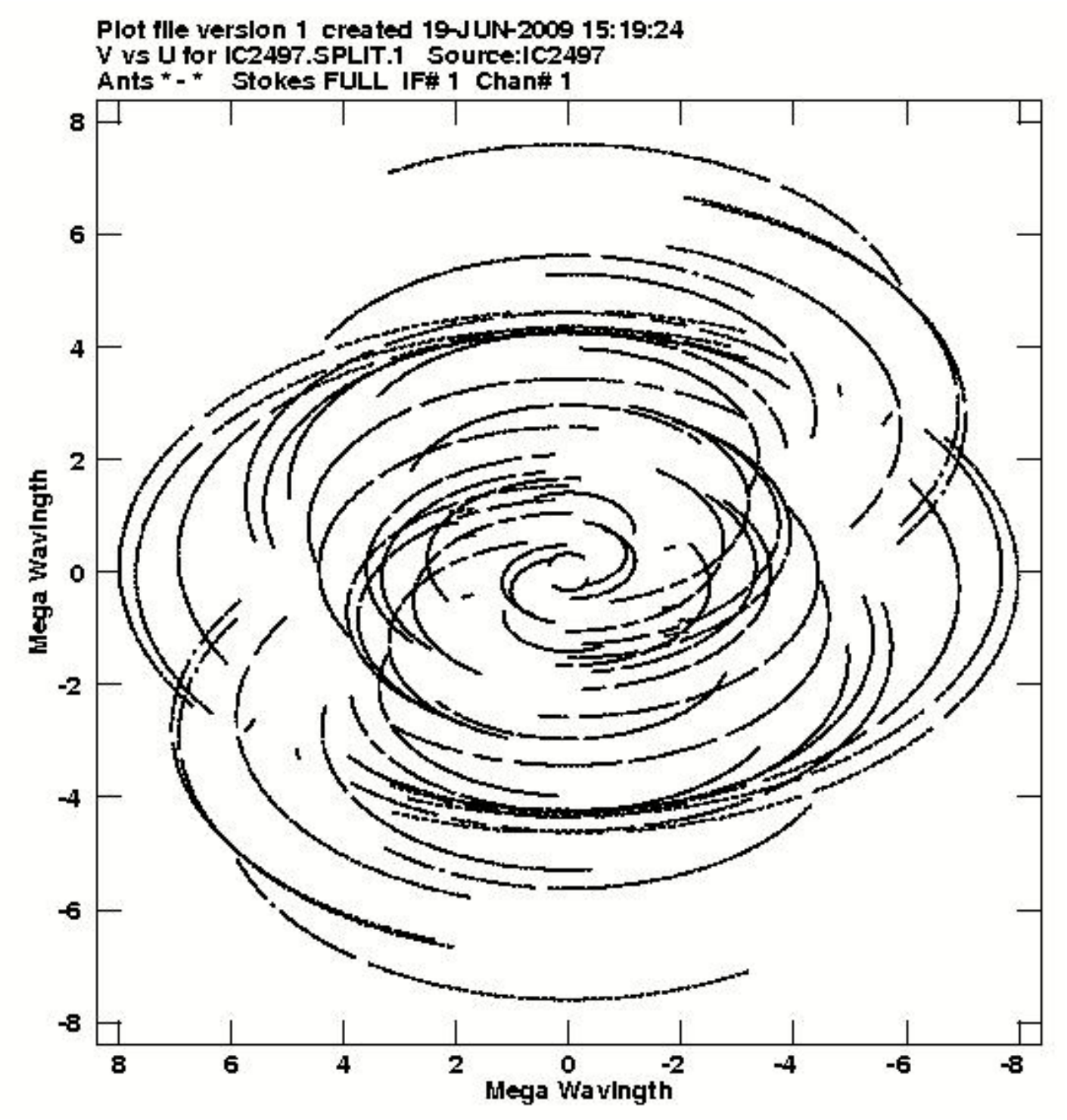} 
\end{minipage}
\\
\\
\end{tabular}
\caption{\textit{Left:} IC~2497 was initially observed with the e-VLBI at $\lambda$ 18~cm, for approximately 2 hours on the 30th September 2008, with 7 telescopes (Wb, Mc, On, Tr, Ef, Jb(MkII) and Da.) \textit{Right:}  The second $\lambda$ 18~cm eVLBI observation, took place on the 19th May 2009, and included 10 telescopes (Ef, Mc, On, Tr, Wb, Cm, Jb (Lovell), Da, De, $\&$ Kn. ). Total observing time was 10 hours. } \label{theLabelOfEntireFigures}
\end{figure}

\begin{figure}

\begin{tabular}{cc}

\begin{minipage}{2.5in}
\centering
\includegraphics[scale=0.35]{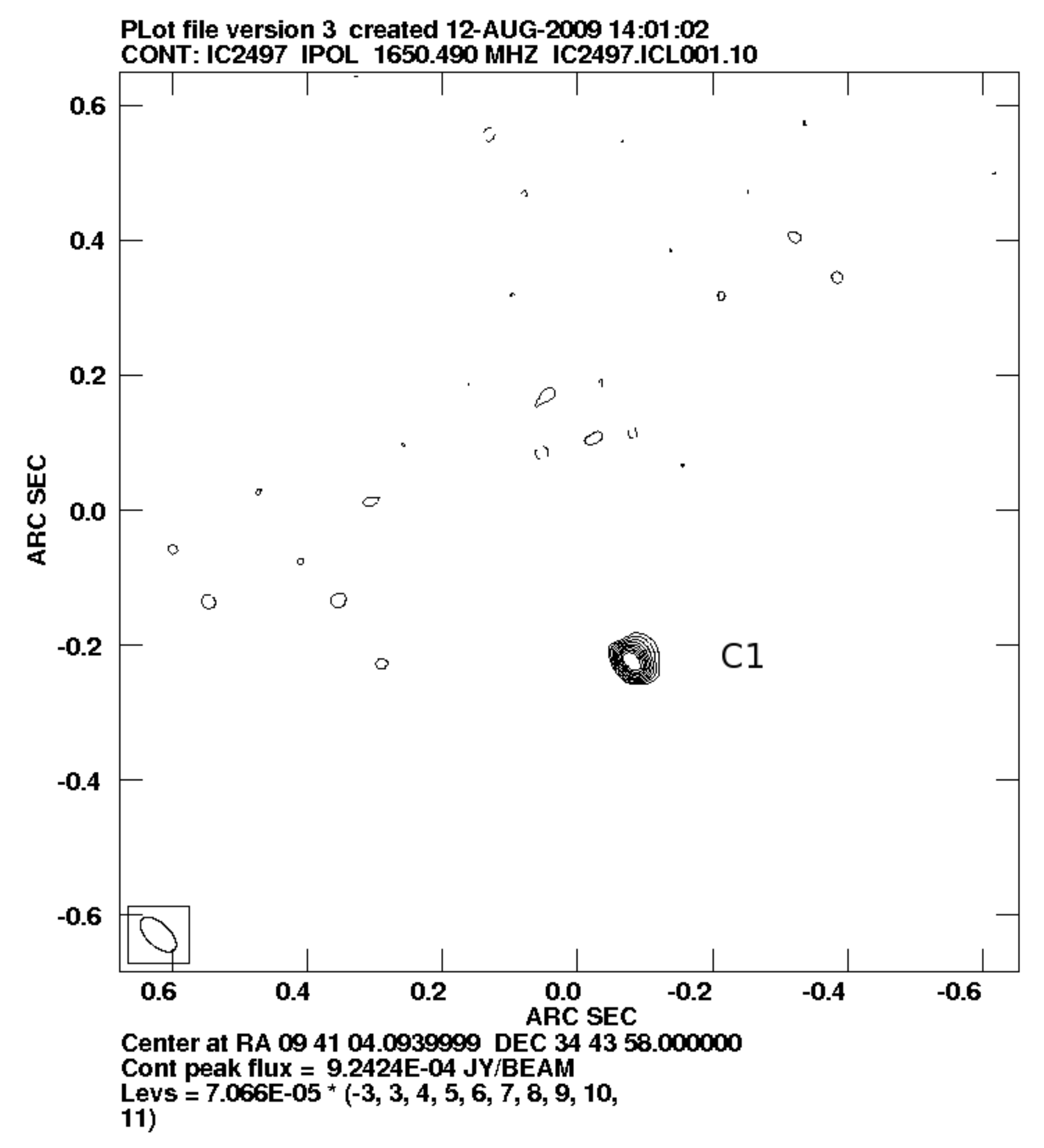} 
\end{minipage}
&
\begin{minipage}{3.0in}
\centering
\includegraphics[scale=0.35]{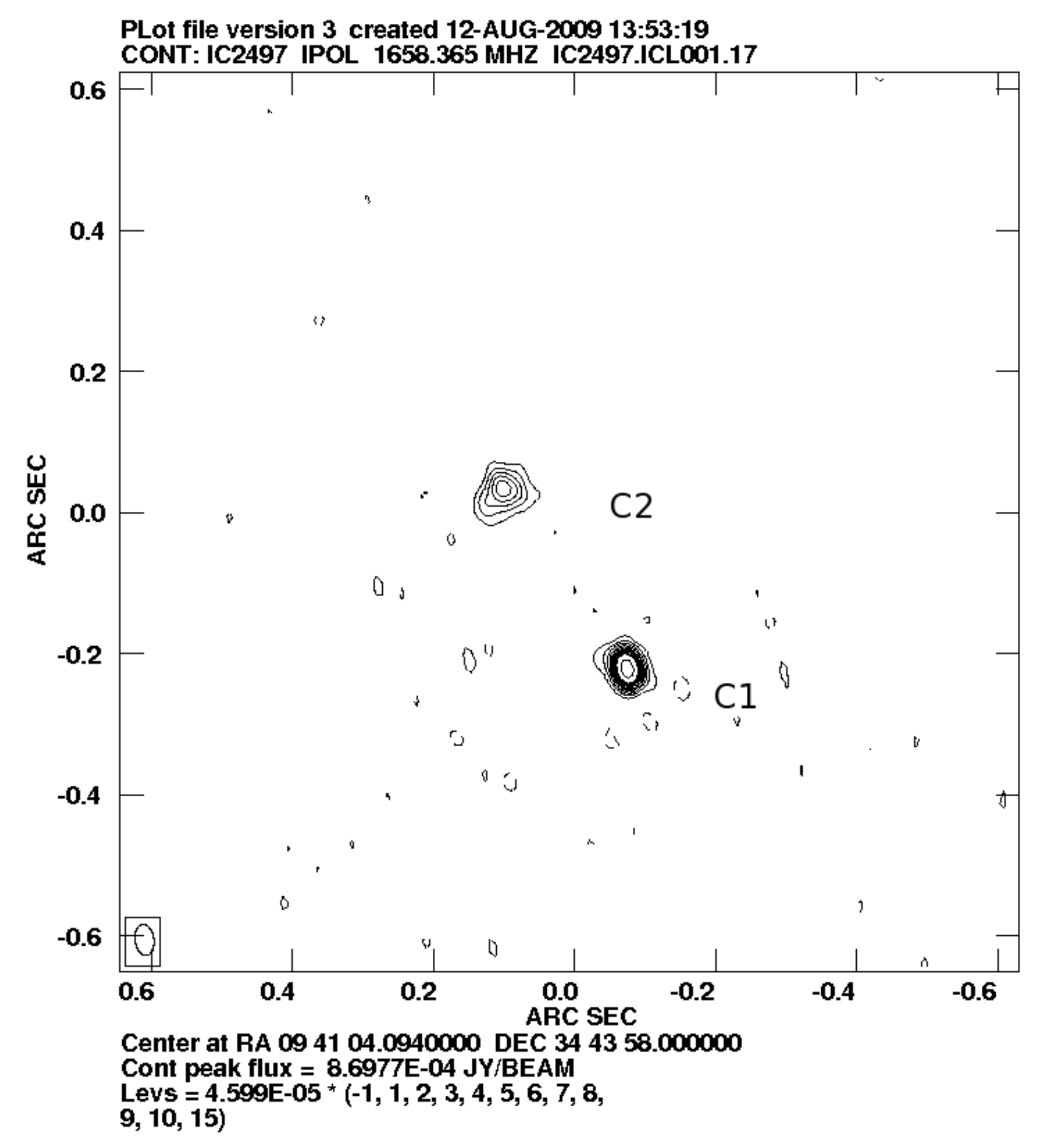} 
\end{minipage}

\\
\end{tabular}
\caption {\textit{Left:} $\lambda$ 18~cm e-VLBI radio map of IC~2497 on 2008 September 2008, with the component C1. The contours are at -3,3,4,5,6,7,8,9,10 $\times$ 0.071 ${\rm mJy}\,{\rm beam}^{-1}$ , with a rms of 0.076 ${\rm mJy}\,{\rm beam}^{-1}$ . The beam size is 58 $\times$ 35 $mas^{2}$ at P.A. = 49.$6^{o}$ . \textit{Right:} $\lambda$ 18~cm e-VLBI radio map of IC~2497 on 2009 May 19th, showing both components, C1 $\&$ C2. The contours are at - 1,1,2,3,4,5,6,7,8,9,10,15 $\times$ 0.046 ${\rm mJy}\,{\rm beam}^{-1}$, with an rms of 0.015 ${\rm mJy}\,{\rm beam}^{-1}$ . The beam size is 45 $\times$ 27 $mas^{2}$ at P.A. = 7.$7^{o}$.} 

\end{figure}


\begin{figure}

\centering
\includegraphics[scale=0.40]{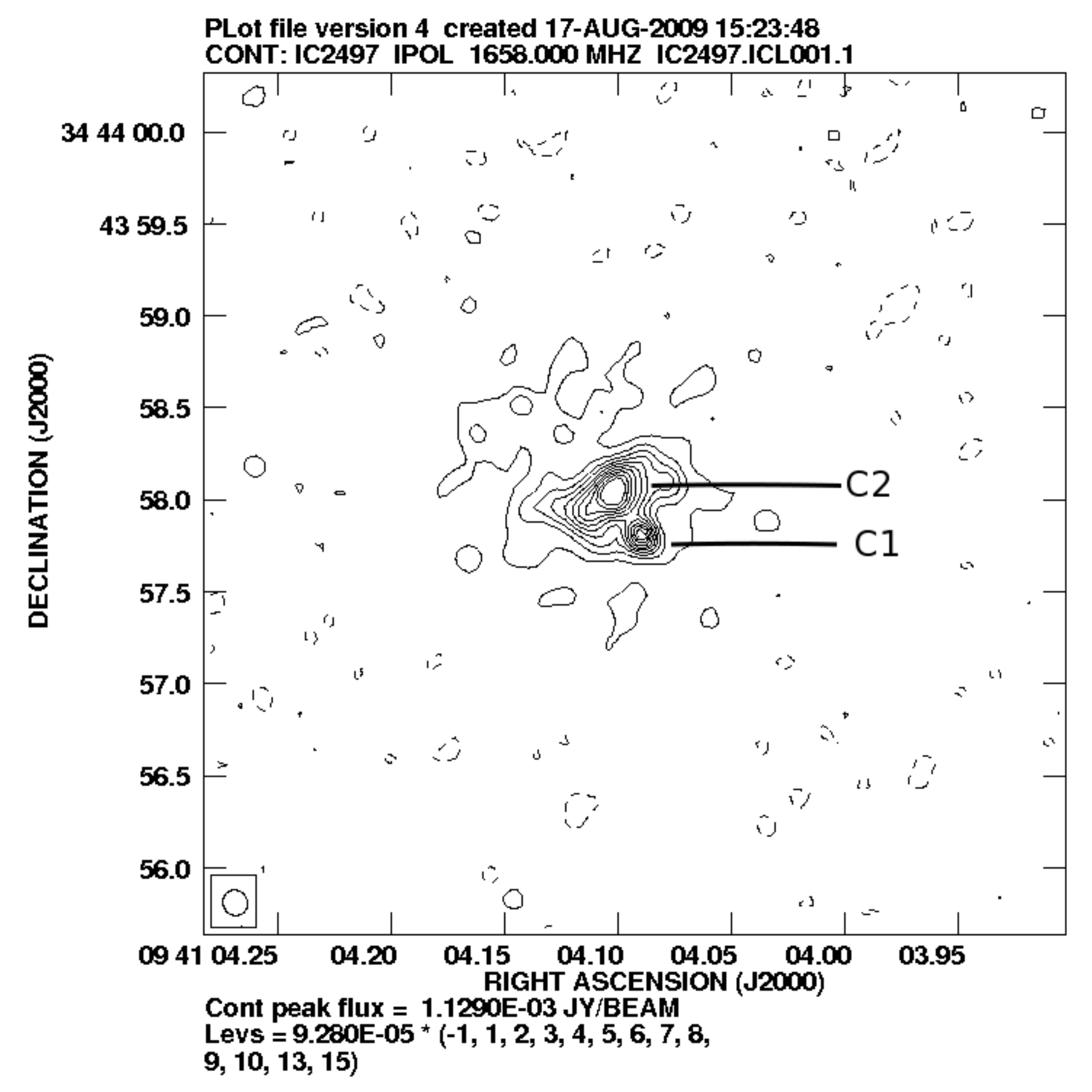} 
\caption{$\lambda$ 18~cm MERLIN radio map of IC~2497 on 2009 February 02, showing both C1 $\&$ C2, embedded within extended emission. The contours are at -1,2,2,3,4,5,6,7,8,9,10,11,12,13,14,15 $\times$ 0.092 ${\rm mJy}\,{\rm beam}^{-1}$ , with an rms 0.037 ${\rm mJy}\,{\rm beam}^{-1}$. The beam size is 181 $\times$ 162 $mas^{2}$ , with a P.A. = 20.$7^{o}$.}

\end{figure}



\section{Discussion}

Our observations further support the hypothesis that IC~2497 contains an obscured AGN at its core \cite{josh_2009}. In this scenario the Voorwerp is illuminated by the AGN after a jet clears a path from the AGN towards the Voorwerp. From our e-VLBI and MERLIN observations we detect 2 compact radio sources (C1 and C2), at the center of IC~2497. From the e-VLBI observations we measure the minimum brightness temperature of both C1 and C2 to be 4 $\times$ 10$^5$ K, and 1.4 $\times$ 10$^5$ K, respectively. The brightness temperature associated with both components, suggest that they are related to AGN activity. 

It is possible we may be observing an AGN core and a hotspot associated with the jet responsible for the Voorwerp. However, from our observations it is very difficult to determine which of the components is the core. Higher frequency observations with MERLIN may be able to differentiate between the two via the spectral index of the components. 

Another possible scenario is that the components C1 and C2 may be radio supernovae (RSNe). In comparison to the radio luminosities of the brightest RSNe from Arp 220 at 18cm \cite{parraetal2007}, we see that C1 is 6 times more luminous while C2 is only 3 times more luminous. From this we cannot exclude the possibility that C2 may be a RSNe.

The non-detection of C2 in the short EVN observation, may not necessarily indicate variability of the source at 18 cm. Although the flux density of C2 is $\sim$ half of C1, most of the flux is in extended emission, which could not be recovered in the short EVN observation. The peak brightness of C2 is 0.236 mJy/beam, which is at the 3 sigma rms noise level of the first EVN observation. Note in the second EVN observation the rms is 0.0015 ${\rm mJy}\,{\rm beam}^{-1}$, which is $\sim$ 5 times lower than the rms of the first EVN observation.

\section*{Acknowledgements}
This research was supported by the EU Framework 6 Marie Curie Early Stage Training programme under contract number MEST-CT-2005-19669 "ESTRELA".
e-VLBI developments in Europe are supported by the EC DG-INFSO funded Communication Network Developments project 'EXPReS', Contract No. 02662.
The European VLBI Network is a joint facility of European, Chinese, South African and other radio astronomy institutes funded by their national research councils.
MERLIN is a National Facility operated by the University of Manchester at Jodrell Bank Observatory on behalf of STFC.

  \end{document}